\documentclass[preprint,5p,twocolumn,times]{elsarticle}
\usepackage{amsmath,amssymb}
\usepackage{bm}
\usepackage{graphicx}

\begin{document}

 \title{Shift manipulation of intrinsic localized mode  \\in ac driven Klein Gordon lattice}
\author[1]{Hirotaka Araki}
\ead{h-araki@dove.kuee.kyoto-u.ac.jp}
\author[1]{Takashi Hikihara} 
\ead{hikihara.takashi.2n@kyoto-u.ac.jp}
\address[1]{Department of Electrical Engineering, Kyoto University, Katsura, Nishikyo, Kyoto 615-8510, Japan.}
\begin{abstract}
Shift manipulation of intrinsic localized mode (ILM) is numerically discussed in an ac driven Klein Gordon lattice. 
Before the manipulation, we introduce the 2-degree of freedom nonlinear system, which is obtained by reducing the lattice. 
In the reduced system, two localized modes correspond to ILMs in the original system. 
These two localized modes are switched by changing a coupling constant adiabatically. 
Based on the result obtained in the reduced system, we numerically show that the shift manipulation of ILM is achieved, like a particle, by using an adiabatic change of coupling constant between the targeted sites in the lattice.
\end{abstract}
\maketitle

\section{Introduction}
Spatially localized and  periodic vibration modes existing in nonlinear discrete systems are called Intrinsic localized modes (ILMs) or discrete breathers (DBs). 
ILMs were theoretically discovered by Sievers and Takeno in 1988\ \cite{sievers1988intrinsic}. 
After the discovery, numerous analytical, numerical, and experimental studies have been reported\ \cite{flach2008discrete}. 
Many researchers have analytically and numerically investigated ILMs from various perspectives such as the existence\ \cite{mackay1994proof, PhysRevE.51.1503}, the stability\ \cite{aubry1997breathers, mackay1998stability} and the movability\ \cite{chen1996breather, aubry1998mobility}. 
ILMs, on the other hand, have been experimentally observed in various physical systems, for example, Josephson-junction array\ \cite{trias2000discrete}, waveguide arrays\ \cite{eisenberg1998discrete}, coupled cantilever array\ \cite{sato2003observation}, and nonlinear transmission line\ \cite{stearrett2007experimental, narahara2020dissipative}. 
Also, molecular dynamics have revealed that ILMs can appear in nanoscale systems\ \cite{savin2008discrete, PhysRevB.81.214306, PhysRevB.84.144303, khadeeva2011discrete, murzaev2017localized}. 
These experimental and numerical studies suggest the phenomenological universality of ILMs and the possibility of applications in both fundamental science and practical engineering. 
Indeed, recently, studies for ILM applications have been reported in energy management in device\ \cite{sato2007management}, energy harvesting\ \cite{jin2014analysis}, and sensors, actuators\ \cite{kimura2008stability, kimura2009coupled, kimura2009capture}, and so on. 
ILMs also have the potential to be used for phonon engineering in nonlinear lattice because ILMs can scatter phonons\ \cite{cretegny19981d, hadipour2020interaction} and affect thermal conductivity\ \cite{tsironis1999dependence}. 

Control of ILMs is smoothly considered as a key to achieve ILM applications as mentioned above. 
It includes generation, annihilation, and spatial position manipulation of ILMs. 
These manipulation has been investigated numerically and experimentally\ \cite{sato2007management, sato2006colloquium}. 
Based on the results, we are going to focus on the spatial position manipulation. 
Sato et al. experimentally and numerically demonstrated a shift of the position of ILMs using a local impurity in nonlinear transmission line and coupled cantilever array\ \cite{sato2007management, sato2004optical}. 
The manipulation is based on the observations that ILMs is attracted or repulsed by the impurity.  
Kimura at el. proposed another method for the spatial position manipulation of ILMs in coupled cantilever array, which is called the capture and release manipulation of the traveling ILMs\ \cite{kimura2009capture}. 
In the method, a nonlinear coupling coefficient is varied at the appropriate time to switching between the pinned stable ILM and the traveling ILM. 
It is based on the stability change due to the saddle-node bifurcation of ILMs.  

Spatial position manipulation of ILMs is also investigated for engineering applications in nonlinear optics. 
In this field, ILMs appear as discrete optical solitons in periodic structures describing arrays of coupled optical waveguides. 
Light propagation in the waveguide arrays is described by nonlinear Schr\"{o}dinger equation in tight binding approximation. 
In the equation, the propagation distance along the waveguide plays the role of time in the dynamic lattice equations. 
Mobility of discrete optical solitons is related to an application for multiport optical switching\ \cite{bang1996exploiting}. 
So, for the unambiguously controlled switching using an appropriate perturbation, there are various proposed methods. 
Rodrigo et al. numerically demonstrate digitized switching\ \cite{vicencio2003controlled, vicencio2004switching}. 
They used a steplike variation of the waveguide coupling for adjusting Peierls-Nabarro (PN) potential appeared from the lattice discreteness\ \cite{kivshar1993peierls}. 
To enhance the mobility of discrete optical solitons, other studies investigate bifurcations of solitons and behavior of PN potential when varying the power of solitons\ \cite{oster2003enhanced, hadvzievski2004power, vicencio2006discrete}. 

We propose a new method for shift manipulation of ILMs in an ac driven Klein Gordon lattice. 
The manipulation corresponds to the moving of an ILM from the site on which the ILM exists to its neighboring site in the specified direction without decaying. 
We specifically target strongly localized ILMs that have a large amplitude and, accordingly, a high frequency relative to plane waves in the phonon band. 
In Klein Gordon lattices, generally, moving ILMs are found only if the amplitude is sufficiently small, which implies that these ILMs have the frequency near the phonon band edge and are weakly localized\ \cite{bang1995high}. 
On the other hand, strongly localized ILMs have low mobility and if they do move, it stops immediately with radiation\ \cite{aubry1998mobility, flach1998discrete}. 
In particular, it is numerically shown that strongly localized ILM is not moved by kicking by small perturbations in Klein Gordon lattice with the quartic potential which is used in this letter\ \cite{chen1996breather}. 

This paper is organized as follows. 
First, we introduce the ac driven Klein Gordon lattice and ILMs which is subject to the manipulation. 
Next, we reduce the lattice to the 2-degree of freedom nonlinear system to avoid the difficulty of the multi-degree of freedom of the lattice. 
In the reduced system, we numerically investigate two coexisting localized modes and an anti-phase mode and implement switching of these localized modes by an adiabatic change of the coupling constant. 
On the basis of this result, finally, we prepare and achieve the shift manipulation of ILM by the adiabatic change of the coupling constant in the original system. 
Furthermore, we investigate conditions to certainly implement the manipulation. 

\section{The model and ILMs}
In this letter, we consider shift manipulation of ILM in ac driven Klein Gordon lattice
The Hamiltonian of Klein Gordon lattice is written as 
\begin{equation}
H=\sum_{n=1}^{N}\left[\frac{\dot{u}_n^2}{2m}+V(u_n)+\frac{C}{2}(u_{n+1}-u_n)^2\right],
\label{eq:energy}
\end{equation} 
where $u_n$ is displacement of the $n$-th site particle from its equilibrium point, $\dot{u}_n$ is its velocity, $m$ is the mass of each particle and $C$ is a coupling constant. 
We take the quartic potential expressed by the following equation
\begin{equation}
V(z)=\frac{z^2}{2}+\frac{z^4}{4}, 
\end{equation}
for the on-site potential with hard-type anharmonicity. 
The investigation of this manipulation for the case of soft-type anharmonicity is summarized in the appendix A. 

The equation of motion of Klein Gordon lattice with dissipation and periodic external driver is written as 
\begin{align}
m\ddot{u}_n+\gamma\dot{u}_n+u_n+u_n^3-C(u_{n-1}-2u_n&+u_{n+1}) \nonumber \\ 
&=(-1)^nF\cos\omega{t}, 
\label{eq:eom}
\end{align}
where $\gamma$ is dissipation coefficient, $F$ is the driving amplitude, and $\omega$ is driver frequency. 
Each particles are periodically driven in reverse phase. 

This model is derived from the coupled cantilever arrays devised by M. Sato and A. J. Sievers. 
For simplicity, however, we neglect the influence of longer than the nearest-neighbor interaction range and the higher-order nonlinearity in the interaction potential. 
We use the reverse-phase driver to sustain ILMs during the manipulation. 
Each parameters are set as follows based on previous researches\ \cite{sato2006colloquium, maniadis2006mechanism}: $m=1,\ C=0.07953,$ and $\gamma=1.534\times10^{-4}$. 
We consider the lattice with the periodic boundary condition consisting of $N=10$ particles. 
We also set the driving amplitude and frequency: $F=0.008$ and $\omega=1.5$ respectively. 

We briefly discuss the phonon band of the lattice. 
As to the case of sufficiently small amplitude vibrations in Eq. (\ref{eq:eom}) with $F=0$ and $\gamma=0$, the nonlinear term can be neglected and we obtain
\begin{equation}
m\ddot{u}_n+u_n-C(u_{n-1}-2u_n+u_{n+1})=0, 
\end{equation}
which is the linearized system of the Klein Gordon lattice.  
Considering plane wave solutions $u_u\sim\exp[i(kn-\omega t)]$ in the limit of an infinite lattice\ ($N\rightarrow\infty$), we obtain
\begin{equation}
\omega^2=1+\frac{2C}{m}(1-\cos k), 
\end{equation}
which is the dispersion relation between the wave number $k$ and the angular frequency $\omega$. 
This equation indicates that the lattice supports the plane waves (phonons) with frequencies ranging from $\omega_{\mathrm{min}}=1$ to $\omega_{\mathrm{max}}=\sqrt{1+4C/m}\simeq1.148$. 
The width of this phonon band is very narrow and the driver frequency is significantly larger than $\omega_{\mathrm{max}}$ which is the top of the phonon band. 

\begin{figure}[htb]
     \begin{center}
    \includegraphics[width=80mm]{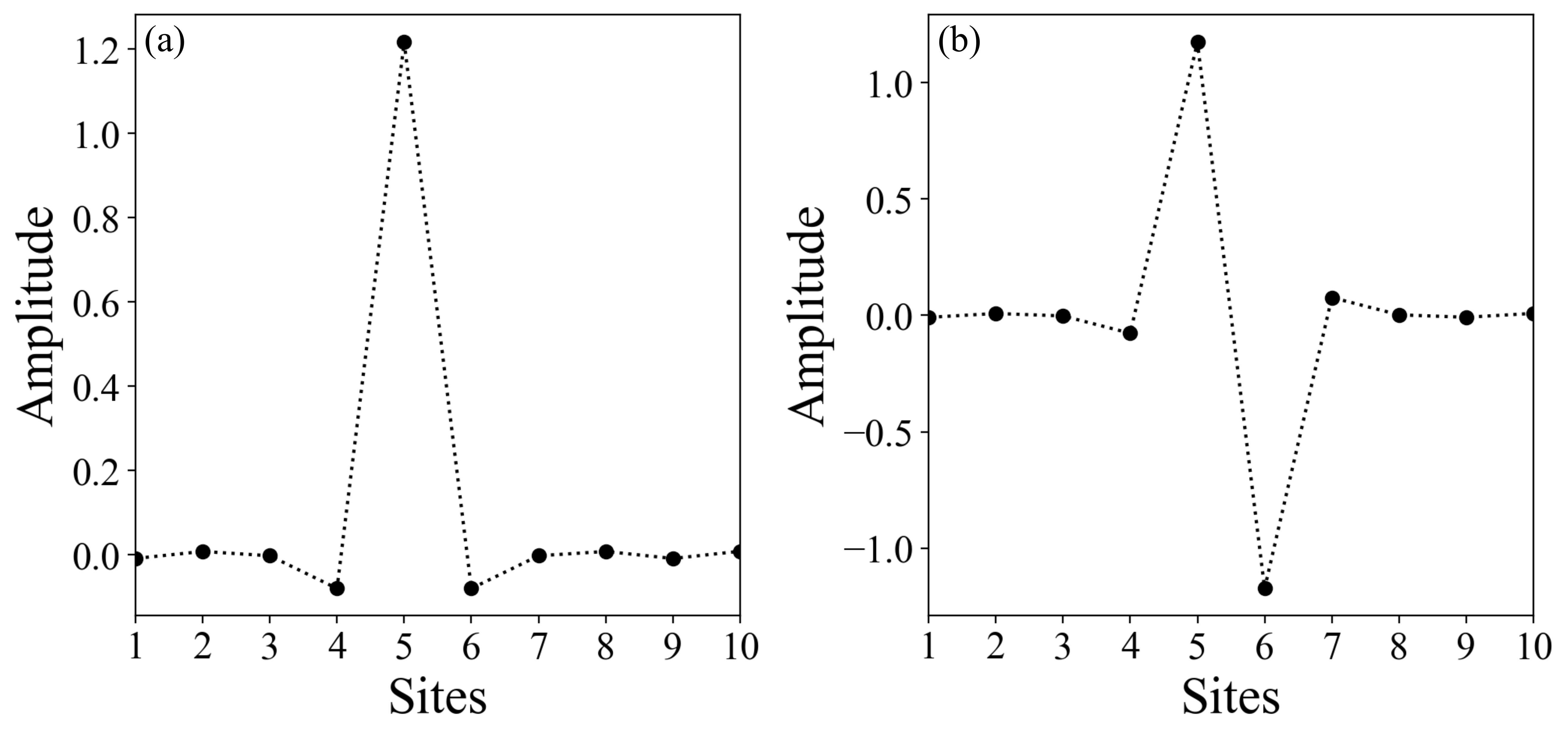}
    \end{center}
       \caption{Amplitude distribution of Sievers-Takeno mode (a) and Page mode (b). Each points indicate the position of particles when they have largest displacements. Each particles oscillate with the same period.}
     \label{fig:one}
\end{figure}

There are two kinds of ILM with different spatial symmetry in this lattice. 
One is Sievers-Takeno mode which is spatially symmetric and centered on sites\ \cite{sievers1988intrinsic}. 
The other is Page mode which is spatially antisymmetric and centered between sites\ \cite{page1990asymptotic}. 
Fig. \ref{fig:one}(a) shows the amplitude distribution of the Sievers-Takeno mode. 
Fig. \ref{fig:one}(b) shows the amplitude distribution of Page mode. 
By using the linear stability analysis method based on Floquet theory, it is confirmed that Sievers-Takeno mode is stable and Page mode is unstable. 
Both of them has almost the same frequency as the driver and is in-phase to it. 
Thus, these ILMs are the strongly localized vibration modes with frequency far away from the phonon band. 
In this letter, we focus Sievers-Takeno mode as the target of shift manipulation. 

\section{The reduced system and NNMs}
At this stage, it seems to be difficult to directly devise a method of the manipulation in this lattice due to its nonlinearity and multi-degree-of-freedom nature. 
So, in the following, we introduce a reduced system of the lattice to consider the shift manipulation of ILM.  
Here, we focus on two sites. 
One is a site where the ILM exists and the other is its neighboring site at the direction to move the ILM. 
The manipulation is a transition of ILM between two adjacent sites. 
The setting makes the influence from outside those two sites neglect. 
This is allowed by the localization of ILM and the sufficiently small coupling constant. 
As a result, we reduce the ac driven Klein Gordon lattice to 2-degrees of freedom system. 
The equation of motion of this 2-degree of freedom nonlinear system is written as 
\begin{align}
&\ddot{u}_{1}+\gamma\dot{u}_{1}+u_{1}+u_{1}^3-C(u_{2}-u_{1})=-F\cos\omega{t},\label{eq:2dfsystem1} \\
&\ddot{u}_{2}+\gamma\dot{u}_{2}+u_{2}+u_{2}^3-C(u_{1}-u_{2})=F\cos\omega{t},
\label{eq:2dfsystem2}
\end{align}
where subscripts are replaced by 1 and 2 for simplicity. 
The reduction allows us to avoid the disturbance to the mode by the structure of the original lattice. 

We investigate the system described by Eq. (6) and (7)  using the concept of nonlinear normalized modes (NNMs)\ \cite{rosenberg1960nonlinear, kerschen2009nonlinear, peeters2009nonlinear}. 
NNMs is proposed by extending the concept of linear normal modes to examine nonlinear vibration systems in which superposition principle and no-interference between modes no longer hold. 
The most simplest definition of NNMs is a vibration in unison of the system. 

We obtain three coexisting NNMs by the shooting method\ \cite{peeters2009nonlinear}. 
A detailed description of this numerical method is given in the appendix B. 
Fig. \ref{fig:two} shows the behavior of each NNM. 
Fig. \ref{fig:two}(a) shows the anti-phase mode which is unstable. 
Fig. \ref{fig:two}(b) shows the localized mode which are the NNM localized at oscillator 1. 
Fig. \ref{fig:two}(c) shows the localized mode which are the NNM localized at oscillator 2. 
These two localized modes are specific to NNMs and stable. 
The anti-phase mode corresponds to unstable Page-mode in the original system. 
Two localized modes correspond to stable Sievers-Takeno modes. 
So, shift manipulation of ILM in the ac driven Klein Gordon lattice corresponds to switching these two coexisting localized modes in the reduced system. 

\begin{figure}[htb]
     \begin{center}
    \includegraphics[width=85mm]{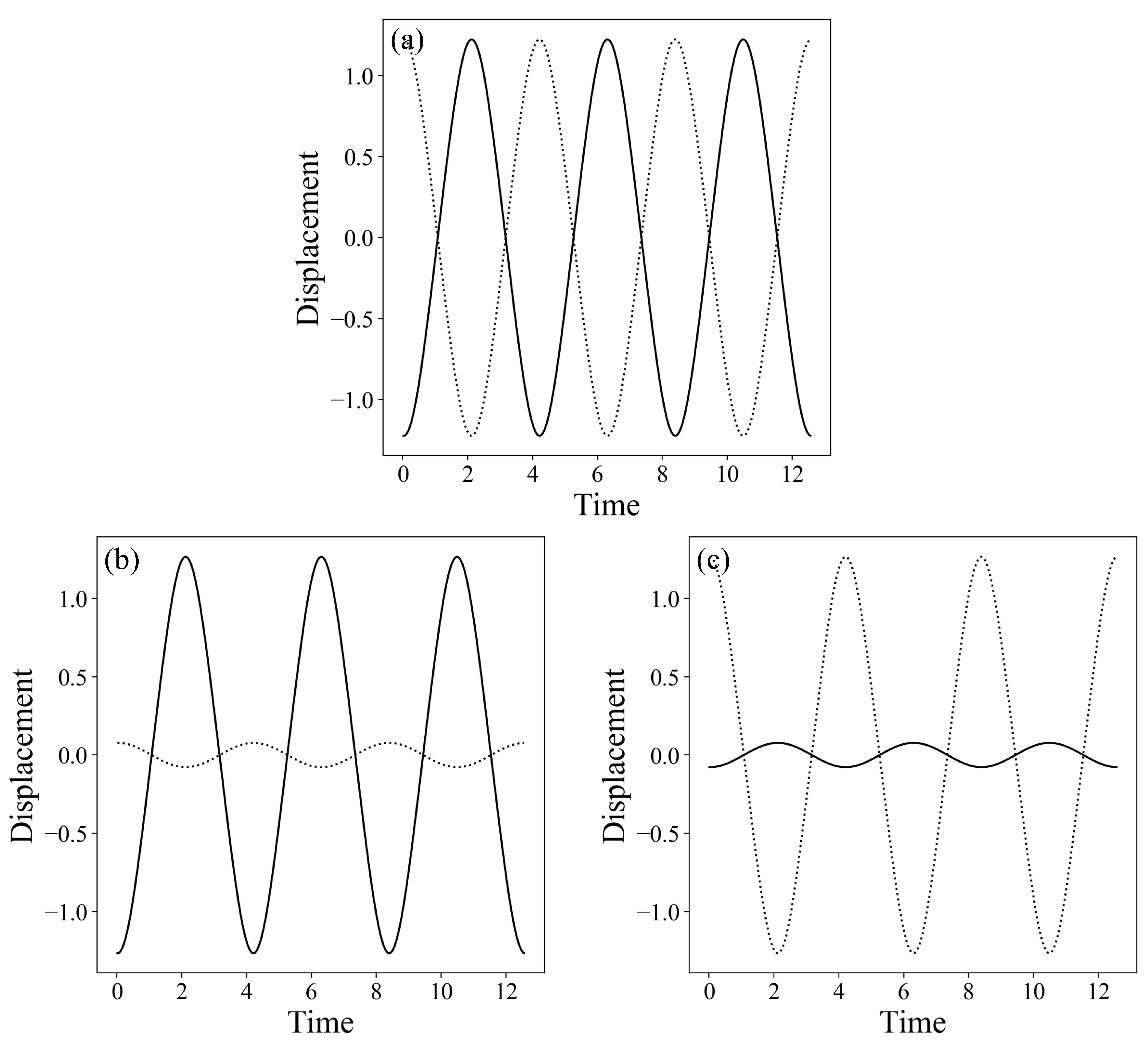}
    \end{center}
       \caption{Time development of three NNMs in the reduced system. Solid lines is displacement of oscillator 1. Dotted lines is displacement of oscillator 2. (a) An anti-phase mode. (b) and (c) Two localized modes.}
     \label{fig:two}
\end{figure}

\section{Switching of two coexisting localized modes}
In this section, we consider switching of two coexisting localized modes in the reduced system. 

\begin{figure}[htb]
     \begin{center}
    \includegraphics[width=80mm]{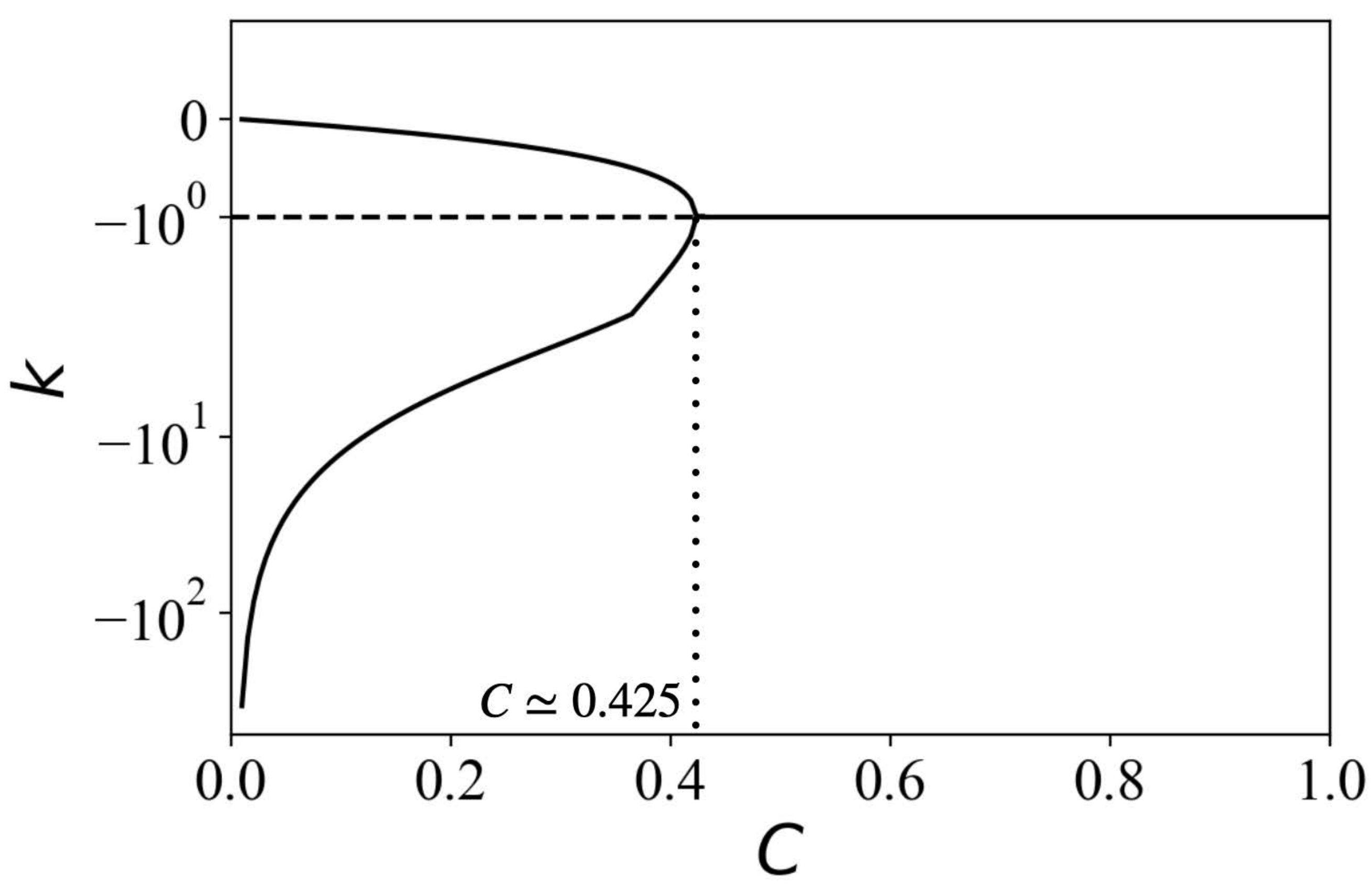}
    \end{center}
       \caption{Relationship between coupling constant and amplitude ratio $k$ in each of the coexisting NNMs. The solid line corresponds to stable NNMs. The dashed line corresponds to unstable NNMs.  } 
\label{fig:three}
\end{figure}

Here, we need to investigate the coupling constant dependency of coexisting NNMs in the system, because the coupling constant is the only parameter that we can tune to achieve the switching. 
We vary the coupling constant and numerically obtain coexisting NNMs at each coupling constant value. 
The NNMs are characterized by the amplitude ratio $k$ defined by $k=r_1/r_2$, where $r_1$ denotes the amplitude of oscillator 1, and $r_2$ the amplitude of oscillator 2. 
Fig. \ref{fig:three} shows amplitude ratio $k$ in each of the coexisting NNMs with respect to the coupling constant $C$. 
The solid line indicates stable NNMs and the dashed line unstable NNMs. 
The straight line at k=1 corresponds to the anti-phase mode. 
The curve above the line corresponds to NNMs localized at oscillator 2. 
The curve below the line corresponds to NNMs localized at oscillator 1. 
When the coupling constant is large, only stable anti-phase mode exists. 
As the value of $C$ is decreased, , the anti-phase mode loses its stability at $C\simeq0.425$ and two coexisting localized modes appear. 
This is a pitchfork bifurcation of NNMs.
 
We consider the switching of two coexisting localized modes on the basis of the bifurcation diagram of NNMs. 
In the diagram, two coexisting localized modes are smoothly connected via the bifurcation point. 
So, we implement the switching two coexisting localized modes through the bifurcation point by tuning the coupling constant. 
This manipulation needs to be achieved adiabatically because a rapid change of the coupling constant destroys NNMs. 
Here, the adiabatic change implies sufficiently slow change compared to the period of driver $T=2\pi/\omega$ which is the characteristic timescale of the system. 
There are infinitely many types of functions that can be used to adiabatically change the coupling constant from the original value to the bifurcation point. 
In this letter, we adopt the following simple way. 
First, we increase the coupling constant linearly from the original value $C=0.07953$ to the bifurcating point $C_{\mathrm{b}}=0.425$. 
This process takes $T_{\mathrm{I}}$. 
Next, we keep the coupling constant at the value $C_{\mathrm{b}}=0.425$ during period $T_{\mathrm{C}}$. 
Lastly, we reduce the coupling constant to the original value during period $T_{\mathrm{R}}$. 
In summary, the coupling constant $C(t)$ during the manipulation is expressed as the following equation, 
\begin{equation}
C(t)=
\begin{cases}
C & (t<t_0), \\  
C+\frac{t-t_0}{T_{\mathrm{I}}}(C_{\mathrm{b}}-C) & (t_0\leq t<t_{\mathrm{I}}),  \\
C_{\mathrm{b}} & (t_{\mathrm{I}}\leq t<t_{\mathrm{C}}),  \\
C_{\mathrm{b}}+\frac{t-t_{\mathrm{C}}}{T_{\mathrm{R}}}(C-C_{\mathrm{b}}) & (t_{\mathrm{C}}\leq t<t_{\mathrm{R}}), \\
C & (t_{\mathrm{R}}\leq t),
\end{cases}
\label{eq:couplingconstant}
\end{equation}
where $t_0$ is a start time of the manipulation, $t_\mathrm{I}=t_0+T_{\mathrm{I}}$, $t_{\mathrm{C}}=t_0+T_{\mathrm{I}}+T_{\mathrm{C}}$ and $t_{\mathrm{R}}=t_0+T_{\mathrm{I}}+T_{\mathrm{C}}+T_{\mathrm{R}}$ which is an end time of the manipulation. 
For changing the coupling constant adiabatically, we need to set the value of $T_{\mathrm{I}}$ and $T_{\mathrm{R}}$ sufficiently larger than $T$. 

\begin{figure*}[htb]
     \begin{center}
    \includegraphics[width=182mm]{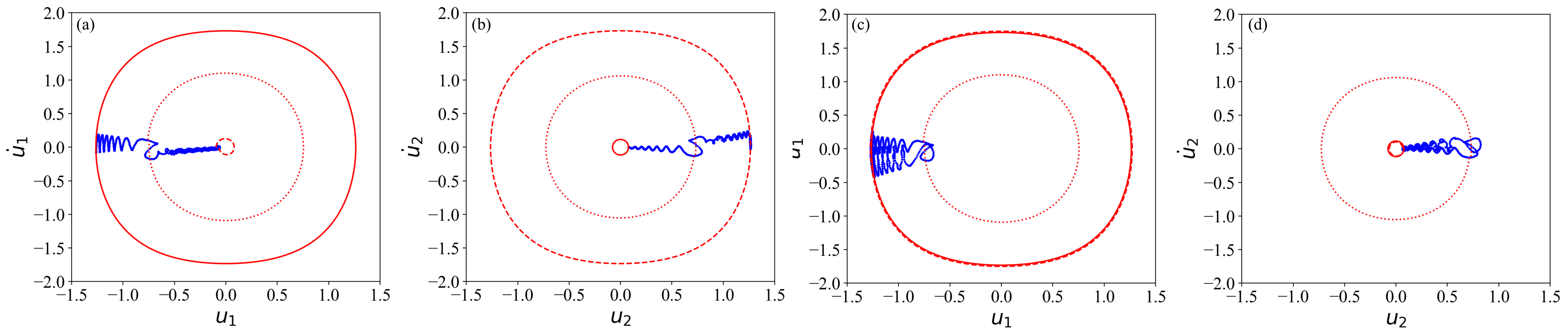}
    \end{center}
       \caption{(a) and (b) Simulation results of the switching when $T_{\mathrm{I}},\ T_{\mathrm{C}}$, and $T_{\mathrm{R}}$ are set at $250T,\ 30T$, and $250T$ respectively. (c) and (d) Simulation results of the switching when $T_{\mathrm{I}},\ T_{\mathrm{C}}$, and $T_{\mathrm{R}}$ are $250T,\ 40T$, and $250T$ respectively. Points (Blue) indicates Poincare map of orbits of each oscillators during the manipulation. Solid line is trajectory for a period of each oscillators before the switching and dotted line is during period $T_\mathrm{C}$, and dashed line is after the switching.}
     \label{fig:five}
\end{figure*}

Here, we switch the mode from NNM localized at oscillator 1 to one at oscillator 2. 
To this goal, we numerically performed the switching with $T_{\mathrm{I}}=250T,\ T_{\mathrm{C}}=30T,\ T_{\mathrm{R}}=250T$. 
Figs. \ref{fig:five}(a) and \ref{fig:five}(b) show results in the phase space of each oscillators. 
Points indicate Poincare maps of orbits of each oscillator during the switching. 
The solid line is each oscillator trajectory for one period before the switching. 
The dotted line is during period $T_\mathrm{C}$. 
The dashed line is after the switching. 
Before the switching, the oscillator 1 has a large amplitude while the oscillator 2 has a small amplitude and each trajectory is closed. 
During period $T_\mathrm{C}$, the oscillators have almost the same amplitude. 
After the switching, the oscillator 1 has a small amplitude while the oscillator 2 has a large amplitude. 
Trajectories of each oscillator during and after the switching (the dotted line and the dashed line) are not closed. 
This is because that linear modes with small amplitudes are excited during the switching. 
The Poincare map points connects trajectories of each oscillator before, during and after the switching. 
We can summarize the result as follows. 
First, during the period $T_\mathrm{I}$, the NNM localized at oscillator 1 (the solid line) gradually approaches the anti-phase mode (the dotted line). 
The anti-phase mode appears during the period $T_{\mathrm{C}}$. 
Finally, during the period $T_\mathrm{R}$, the anti-phase mode approaches NNM localized at oscillator 2 (the dashed line). 
Thus, under this condition, the switching is achieved between the coexisting localized modes. 

We then performed the switching with $T_{\mathrm{I}}=250T,\ T_{\mathrm{C}}=40T,\ T_{\mathrm{R}}=250T$. 
Figs. \ref{fig:five}(c) and \ref{fig:five}(d) show results. 
Points indicate Poincare maps and lines each oscillator trajectory for one period before, during, and after the switching. 
Before the switching, the oscillator 1 has a large amplitude while the oscillator 2 has a small amplitude and each trajectory is closed. 
During period $T_\mathrm{C}$, the oscillators have almost the same amplitude. 
After the switching, the oscillator 1 has a large amplitude again while the oscillator 2 has a small amplitude. 
Trajectories of each oscillator before and after the switching almost overlap (the solid line and the dashed line). 
Trajectories of each oscillator during and after the switching (the dotted line and the dashed line) are not closed for the same reason as in the previous case. 
We can summarize the result as follows. 
First, during the period $T_\mathrm{I}$, the NNM localized at oscillator 1 (the solid line) gradually approaches the anti-phase mode (the dotted line). 
The anti-phase mode appears during the period $T_{\mathrm{C}}$. 
Finally, during the period $T_\mathrm{R}$, the anti-phase mode approaches the original state (the dashed line). 
Therefore, under this condition, we will not achieve the switching of coexisting localized modes. 
From these results, the result of this switching depends on $T_{\mathrm{C}}$. 
Further numerical investigations reveal that the result of the manipulation depends not only $T_{\mathrm{C}}$, but also other factors in the process such as $T_{\mathrm{I}}$ and $T_{\mathrm{R}}$. 
To be more precise, the result of the switching depends on how to change the coupling constant and the state at the time of starting the manipulation. 

\section{Shift manipulation of ILM}
At last, we consider the shift manipulation of ILM in the ac driven Klein Gordon lattice. 
Based on the results obtained in the reduced system, we expect that the manipulation is achieved by the adiabatic change of a coupling constant between two sites at which ILM originally exists and the neighbor site of the direction to move the ILM. 
Here, we handle an ILM at the 5th site to the 6th site. 
So, we change a coupling constant between the 5th site and the 6th site adiabatically as in the case of switching coexisting localized modes. 
We can prepare the ILM by the shooting method\ \cite{marin1996breathers}. 
This numerical method id described in the appendix B. 

Fig. \ref{fig:seven}(a) shows the behavior of the ILM during the manipulation when $T_{\mathrm{I}},\ T_{\mathrm{C}}$, and $T_{\mathrm{R}}$ are set at $250T,\ 20T$, and $250T$, respectively. 
The heat map indicates the time development of energy at each site. 
The dark region indicates a high energy state. 
Before the manipulation, most of the energy is concentrated at the 5th site, which corresponds to the ILM at the 5th site. 
As we linearly increase the coupling constant to get closer to the bifurcation point during the period $T_{\mathrm{I}}$, the energy of the 5th site gradually excludes into the 6th site. 
When keeping the coupling constant during the period $T_{\mathrm{C}}$, most of the energy is almost equally distributed at both of the 5th and the 6th sites. 
During this period, the amplitude distribution looks like the Page mode. 
However, this is not the Page mode because the lattice is not homogeneous during this period. 
As we linearly reduce the coupling constant to return to the original value during the period $T_{\mathrm{R}}$, the energy starts to gather in the 6th site. 
After the manipulation, most of the energy is concentrated at the 6th site.
As a result, the manipulation succeeds. 

Fig. \ref{fig:seven}(b) shows the behavior of the ILM during the manipulation when $T_{\mathrm{I}},\ T_{\mathrm{C}}$, and $T_{\mathrm{R}}$ are set at $250T,\ 30T$, and $250T$ respectively. 
Before the period $T_{\mathrm{R}}$, this figure shows the same behavior when $T_{\mathrm{C}}$ is set at $20T$. 
However, after the manipulation, the ILM returns to the 5th site.
It shows the failure of the manipulation in this case. 
Thus, the shift manipulation of ILM depends on $T_{\mathrm{C}}$ as in the case of the switching of two coexisting localized modes. 
It consequently seems that the manipulation depends on how to change the coupling constant during the manipulation and the state at the time of starting the manipulation. 

We then investigate the dependency on the process and the initial state. 
These two dependencies are inherently inseparable in this matter. 
Therefore, we first investigate the dependency on the process and, based on the result, we approach the dependency on the initial state.   

\begin{figure*}[htb]
     \begin{center}
    \includegraphics[width=175mm]{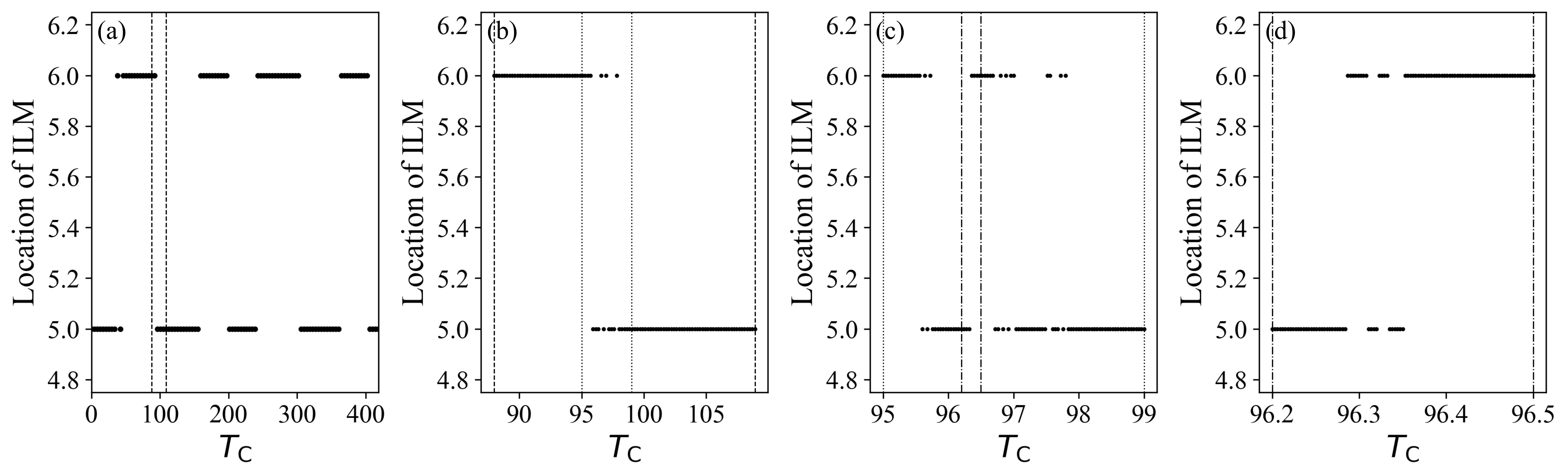}
    \end{center}
       \caption{$T_{\mathrm{C}}$ dependency of the shift manipulation of ILM. The horizontal axis indicates $T_{\mathrm{C}}$. The vertical axis indicates the location of the ILM after the shift manipulation. (a) Results when $T_{\mathrm{C}}$ is varied from $0$ to $100T$ every $T$. (b) Results when $T_{\mathrm{C}}$ is varied from $21T\ (\simeq 88)$ to $26T\ (\simeq 109)$ every $T/20\ (\simeq 0.21)$. (c) Results when $T_{\mathrm{C}}$ is varied from $95$ to $99$ every $1/25$. (d) Results when $T_{\mathrm{C}}$ is varied from $96.2$ to $96.5$ every $3/1000$.} 
        \label{fig:eight}
\end{figure*}

In this letter, the manipulation process is expressed by Eq. (\ref{eq:couplingconstant}), which is characterized by $T_{\mathrm{I}},\ T_{\mathrm{C}}$, and $T_{\mathrm{R}}$ (We do not consider $C_\mathrm{b}$ because it is set at the value of the bifurcation point). 
Among these, we set the values of $T_{\mathrm{I}}$ and $T_{\mathrm{R}}$ based on the adiabatic condition. 
Therefore, we investigate $T_{\mathrm{C}}$ dependency of the manipulation with $T_{\mathrm{I}}$ and $T_{\mathrm{R}}$ fixed at $250T$ respectively. 
To that end, we repeatedly implement the shift simulation varying $T_{\mathrm{C}}$ every $T$ from $T_{\mathrm{C}}=0$ to $T_{\mathrm{C}}=100T$. 
Fig. \ref{fig:eight}(a) shows the location of the ILM after the manipulation at each $T_{\mathrm{C}}$. 
There are two different intervals for the value of $T_{\mathrm{C}}$: the manipulation succeeds on one type of interval and fails on the other.  
These two different intervals appear alternately.
Next, in order to examine boundaries between these two different intervals, we zoom in the boundary closer and closer.  
Fig. \ref{fig:eight}(b) is an enlargement of the section enclosed by the dashed line in Fig. \ref{fig:eight}(a), which is obtained by simulations with $T_{\mathrm{C}}$ varied from $T_{\mathrm{C}}=21T\ (\simeq 88)$ to $T_{\mathrm{C}}=26T\ (\simeq 109)$ every $T/20\ (\simeq 0.21)$. 
Fig. \ref{fig:eight}(c) is an enlargement of the section enclosed by the dotted line in Fig. \ref{fig:eight}(b), which is obtained by simulations with $T_{\mathrm{C}}$ varied from $T_{\mathrm{C}}=95$ to $T_{\mathrm{C}}=99$ every $1/25$.  
Fig. \ref{fig:eight}(d) is an enlargement of the section enclosed by the dashed-dotted line in Fig. \ref{fig:eight}(c), which is obtained by simulations with $T_{\mathrm{C}}$ varied from $T_{\mathrm{C}}=96.2$ to $T_{\mathrm{C}}=96.5$ every $3/1000$. 
These figures show that there are fine nested structures regarding the success or failure of the manipulation near the boundaries. 

\begin{figure}[htb]
     \begin{center}
    \includegraphics[width=80mm]{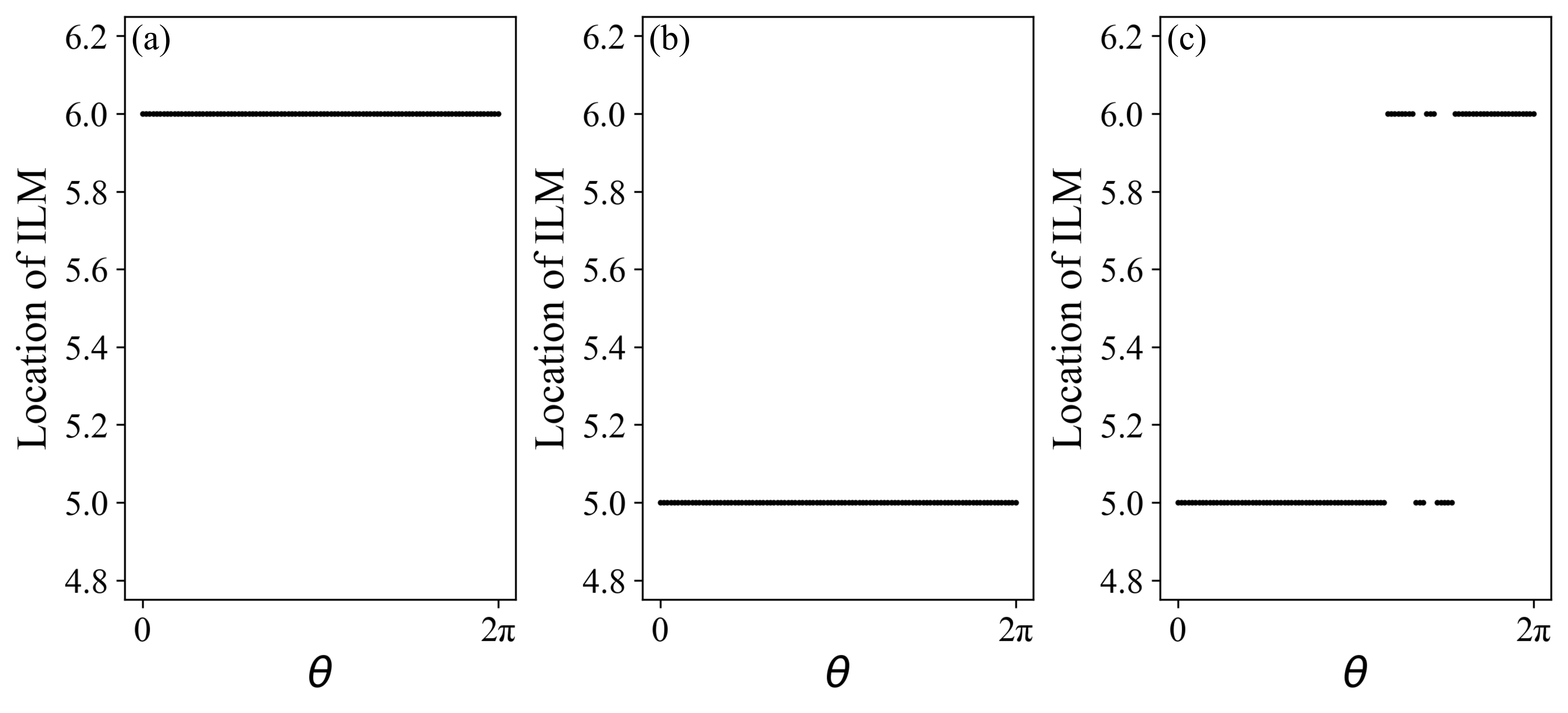}
    \end{center}
       \caption{The initial phase $\theta$ dependency of the shift manipulation of the ILM. The horizontal axis indicates $\theta$. The vertical axis indicates the location of the ILM after the shift manipulation. (a) Results when $T_{\mathrm{C}}=20T\simeq83.73$. (b) Results when $T_{\mathrm{C}}=30T\simeq125.6$. (c) Results when $T_{\mathrm{C}}=96$.}
     \label{fig:fifty}
\end{figure}

We then investigate the dependency on the initial state when starting the manipulation. 
Here, we consider the initial state as the initial phase $\theta$ of the ILM. 
Based on the $T_{\mathrm{C}}$ dependency of the manipulation, we consider the following three cases. 
The first is when the value of $T_{\mathrm{C}}$ is included in an interval on which the manipulation succeeds. 
The second is when the value of $T_{\mathrm{C}}$ is included in an interval on which the manipulation fails.
The third is when the value of $T_{\mathrm{C}}$ is near a boundary between these intervals.
For each case, we vary $\theta$ from $0$ to $2\pi$ every $\pi/50$. 
Fig, \ref{fig:fifty}(a) shows the initial phase $\theta$ dependency for the first case ($T_{\mathrm{C}}=20T\simeq83.73$), fig, \ref{fig:fifty}(b) for the second case ($T_{\mathrm{C}}=30T\simeq125.6$) and fig, \ref{fig:fifty}(c) for the third case ($T_{\mathrm{C}}=96$).
When the value of $T_{\mathrm{C}}$ is included in the two different intervals, the initial phase $\theta$ of the ILM has no influence on the result of the manipulation. 
However, when the value of $T_{\mathrm{C}}$ is near the boundaries, the result of the manipulation is unstable with respect to the value of the initial phase $\theta$. 
Consequently, for the accurate manipulation of ILM, we need to set the value of $T_{\mathrm{C}}$ appropriately. 

\begin{figure}[htb]
     \begin{center}
    \includegraphics[width=88mm]{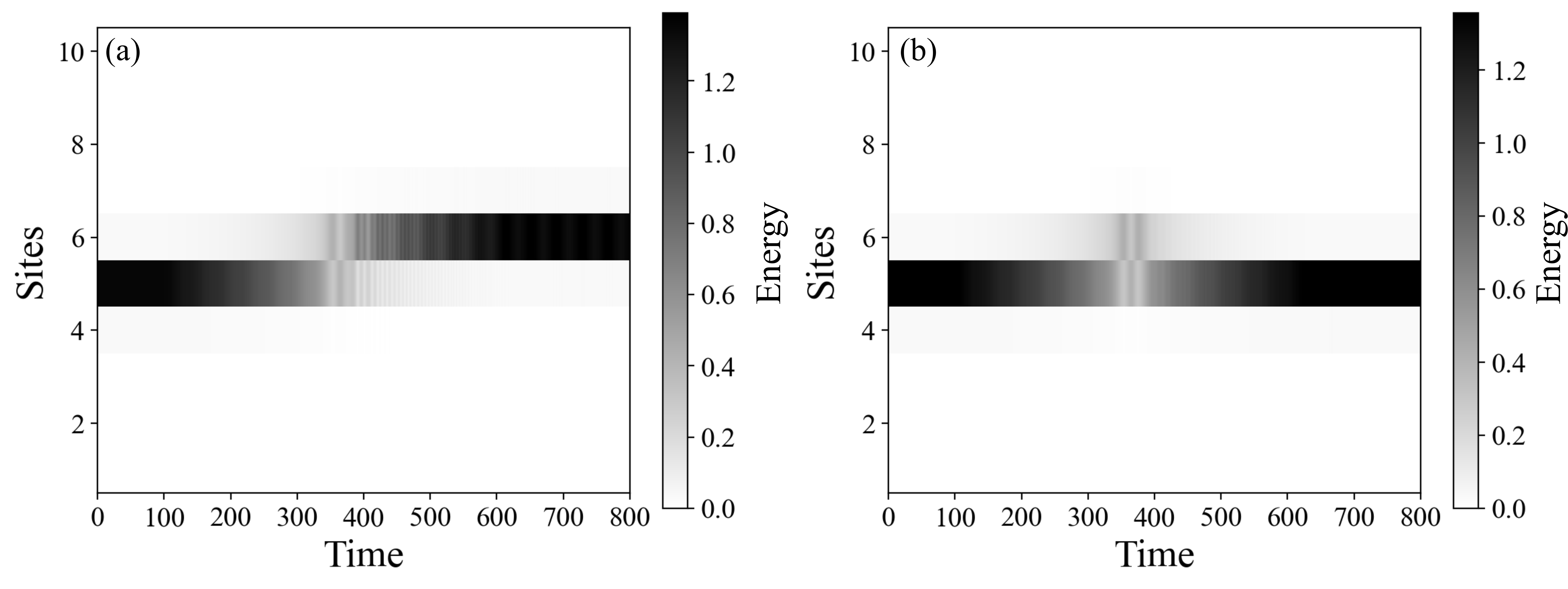}
    \end{center}
       \caption{Time development of energy of each sites during the shift manipulation of ILM. The unit time on the horizontal axis is $T$ which is the period of driver. (a) Simulation results when $T_{\mathrm{I}},\ T_{\mathrm{C}}$, and $T_{\mathrm{R}}$ are set at $250T,\ 20T$, and $250T$ respectively. (b) Simulation results when $T_{\mathrm{I}},\ T_{\mathrm{C}}$, and $T_{\mathrm{R}}$ are set at $250T,\ 30T$, and $250T$ respectively. }
     \label{fig:seven}
\end{figure}

\section{Conclusion}
In this letter, we proposed the shift manipulation of ILM and numerically demonstrated it in ac driven Klein Gordon lattice. 
We specifically targeted the strongly localized ILM for the manipulation.  

At first, we introduced the reduced system of the lattice, which is obtained by neglecting the influence from outside two sites which are the subject of the manipulation. 
There are three NNMs in this reduced system; two stable localized modes corresponding to the stable ILM and an unstable anti-phase mode corresponding to the unstable Page mode. 
We investigated these NNMs' dependency on the coupling constant and found that pitchfork bifurcation of NNMs occurs. 
On the basis of the bifurcation diagram, we implemented the switching of two coexisting localized modes by changing the coupling constant adiabatically. 
We also revealed that the result of the switching depends on its process. 

Lastly, we confirmed that the shift manipulation of ILM is achieved by the adiabatic change of coupling constant between two targeted sites as is the case in the reduced system. 
Furthermore, we investigate the dependency of the manipulation on its process and revealed that there are two different intervals for the value of $T_{\mathrm{C}}$: the manipulation succeeds in one interval and fails in the other. 
We found fine nested structures near boundaries between these intervals. 
These structures seem to be interesting and need to be examined in more detail as to their origin. 
Due to these structures, the result of the manipulation sensitively depends on the initial phase of the ILM. 
Consequently, we have to set the value of $T_{\mathrm{C}}$ appropriately for the accurate manipulation.  

Additionally, in the appendix A, we showed that this manipulation is also achieved in Klein Gordon lattice with soft-type anharmonicity. 
So, it seems that this manipulation method is practicable in a wide range of Klein Gordon lattices. 
We expect that this manipulation is able to be used in other nonlinear systems not included in this paper by following the same schemes as in this paper. 


\bibliography{ref}

\begin{thebibliography}{10}
\expandafter\ifx\csname url\endcsname\relax
  \def\url#1{\texttt{#1}}\fi
\expandafter\ifx\csname urlprefix\endcsname\relax\def\urlprefix{URL }\fi
\expandafter\ifx\csname href\endcsname\relax
  \def\href#1#2{#2} \def\path#1{#1}\fi

\bibitem{sievers1988intrinsic}
A.~J. Sievers, S.~Takeno,
  \href{https://link.aps.org/doi/10.1103/PhysRevLett.61.970}{Intrinsic
  localized modes in anharmonic crystals}, Phys. Rev. Lett. 61~(8) (1988)
  970--973.
\newblock \href {https://doi.org/10.1103/PhysRevLett.61.970}
  {\path{doi:10.1103/PhysRevLett.61.970}}.
\newline\urlprefix\url{https://link.aps.org/doi/10.1103/PhysRevLett.61.970}

\bibitem{flach2008discrete}
S.~Flach, A.~V. Gorbach,
  \href{https://doi.org/10.1016/j.physrep.2008.05.002}{Discrete
  breathers—advances in theory and applications}, Physics Reports 467~(1-3)
  (2008) 1--116.
\newblock \href {https://doi.org/10.1016/j.physrep.2008.05.002}
  {\path{doi:10.1016/j.physrep.2008.05.002}}.
\newline\urlprefix\url{https://doi.org/10.1016/j.physrep.2008.05.002}

\bibitem{mackay1994proof}
R.~S. MacKay, S.~Aubry, \href{https://doi.org/10.1088/0951-7715/7/6/006}{Proof
  of existence of breathers for time-reversible or hamiltonian networks of
  weakly coupled oscillators}, Nonlinearity 7~(6) (1994) 1623--1643.
\newblock \href {https://doi.org/10.1088/0951-7715/7/6/006}
  {\path{doi:10.1088/0951-7715/7/6/006}}.
\newline\urlprefix\url{https://doi.org/10.1088/0951-7715/7/6/006}

\bibitem{PhysRevE.51.1503}
S.~Flach, \href{https://link.aps.org/doi/10.1103/PhysRevE.51.1503}{Existence of
  localized excitations in nonlinear hamiltonian lattices}, Phys. Rev. E 51
  (1995) 1503--1507.
\newblock \href {https://doi.org/10.1103/PhysRevE.51.1503}
  {\path{doi:10.1103/PhysRevE.51.1503}}.
\newline\urlprefix\url{https://link.aps.org/doi/10.1103/PhysRevE.51.1503}

\bibitem{aubry1997breathers}
S.~Aubry, \href{https://doi.org/10.1016/S0167-2789(96)00261-8}{Breathers in
  nonlinear lattices: Existence, linear stability and quantization}, Physica D:
  Nonlinear Phenomena 103~(1-4) (1997) 201--250.
\newblock \href {https://doi.org/10.1016/S0167-2789(96)00261-8}
  {\path{doi:10.1016/S0167-2789(96)00261-8}}.
\newline\urlprefix\url{https://doi.org/10.1016/S0167-2789(96)00261-8}

\bibitem{mackay1998stability}
R.~S. MacKay, J.~A. Sepulchre,
  \href{https://doi.org/10.1016/S0167-2789(98)00073-6}{Stability of discrete
  breathers}, Physica D: Nonlinear Phenomena 119~(1-2) (1998) 148--162.
\newblock \href {https://doi.org/10.1016/S0167-2789(98)00073-6}
  {\path{doi:10.1016/S0167-2789(98)00073-6}}.
\newline\urlprefix\url{https://doi.org/10.1016/S0167-2789(98)00073-6}

\bibitem{chen1996breather}
D.~Chen, S.~Aubry, G.~P. Tsironis,
  \href{https://link.aps.org/doi/10.1103/PhysRevLett.77.4776}{Breather mobility
  in discrete $\varphi^4$ nonlinear lattices}, Physical Review Letters 77~(23)
  (1996) 4776--4779.
\newblock \href {https://doi.org/10.1103/PhysRevLett.77.4776}
  {\path{doi:10.1103/PhysRevLett.77.4776}}.
\newline\urlprefix\url{https://link.aps.org/doi/10.1103/PhysRevLett.77.4776}

\bibitem{aubry1998mobility}
S.~Aubry, T.~Cretegny,
  \href{https://doi.org/10.1016/S0167-2789(98)00062-1}{Mobility and reactivity
  of discrete breathers}, Physica D: Nonlinear Phenomena 119~(1-2) (1998)
  34--46.
\newblock \href {https://doi.org/10.1016/S0167-2789(98)00062-1}
  {\path{doi:10.1016/S0167-2789(98)00062-1}}.
\newline\urlprefix\url{https://doi.org/10.1016/S0167-2789(98)00062-1}

\bibitem{trias2000discrete}
E.~Trias, J.~J. Mazo, T.~P. Orlando,
  \href{https://link.aps.org/doi/10.1103/PhysRevLett.84.741}{Discrete breathers
  in nonlinear lattices: Experimental detection in a josephson array}, Physical
  Review Letters 84~(4) (2000) 741--744.
\newblock \href {https://doi.org/10.1103/PhysRevLett.84.741}
  {\path{doi:10.1103/PhysRevLett.84.741}}.
\newline\urlprefix\url{https://link.aps.org/doi/10.1103/PhysRevLett.84.741}

\bibitem{eisenberg1998discrete}
H.~S. Eisenberg, Y.~Silberberg, R.~Morandotti, A.~R. Boyd, J.~S. Aitchison,
  \href{https://link.aps.org/doi/10.1103/PhysRevLett.81.3383}{Discrete spatial
  optical solitons in waveguide arrays}, Physical Review Letters 81~(16) (1998)
  3383--3386.
\newblock \href {https://doi.org/10.1103/PhysRevLett.81.3383}
  {\path{doi:10.1103/PhysRevLett.81.3383}}.
\newline\urlprefix\url{https://link.aps.org/doi/10.1103/PhysRevLett.81.3383}

\bibitem{sato2003observation}
M.~Sato, B.~E. Hubbard, A.~J. Sievers, B.~Ilic, D.~A. Czaplewski, H.~G.
  Craighead,
  \href{https://link.aps.org/doi/10.1103/PhysRevLett.90.044102}{Observation of
  locked intrinsic localized vibrational modes in a micromechanical oscillator
  array}, Physical Review Letters 90~(4) (2003) 044102.
\newblock \href {https://doi.org/10.1103/PhysRevLett.90.044102}
  {\path{doi:10.1103/PhysRevLett.90.044102}}.
\newline\urlprefix\url{https://link.aps.org/doi/10.1103/PhysRevLett.90.044102}

\bibitem{stearrett2007experimental}
R.~Stearrett, L.~Q. English,
  \href{https://doi.org/10.1088/0022-3727/40/17/058}{Experimental generation of
  intrinsic localized modes in a discrete electrical transmission line},
  Journal of Physics D: Applied Physics 40~(17) (2007) 5394.
\newblock \href {https://doi.org/10.1088/0022-3727/40/17/058}
  {\path{doi:10.1088/0022-3727/40/17/058}}.
\newline\urlprefix\url{https://doi.org/10.1088/0022-3727/40/17/058}

\bibitem{narahara2020dissipative}
K.~Narahara, \href{https://doi.org/10.7566/JPSJ.89.074005}{Dissipative discrete
  breathers in series-connected tunnel diode oscillator lattice}, Journal of
  the Physical Society of Japan 89~(7) (2020) 074005.
\newblock \href {https://doi.org/10.7566/JPSJ.89.074005}
  {\path{doi:10.7566/JPSJ.89.074005}}.
\newline\urlprefix\url{https://doi.org/10.7566/JPSJ.89.074005}

\bibitem{savin2008discrete}
A.~V. Savin, Y.~S. Kivshar,
  \href{https://doi.org/10.1209/0295-5075/82/66002}{Discrete breathers in
  carbon nanotubes}, EPL (Europhysics Letters) 82~(6) (2008) 66002.
\newblock \href {https://doi.org/10.1209/0295-5075/82/66002}
  {\path{doi:10.1209/0295-5075/82/66002}}.
\newline\urlprefix\url{https://doi.org/10.1209/0295-5075/82/66002}

\bibitem{PhysRevB.81.214306}
L.~Z. Khadeeva, S.~V. Dmitriev,
  \href{https://link.aps.org/doi/10.1103/PhysRevB.81.214306}{Discrete breathers
  in crystals with nacl structure}, Phys. Rev. B 81 (2010) 214306.
\newblock \href {https://doi.org/10.1103/PhysRevB.81.214306}
  {\path{doi:10.1103/PhysRevB.81.214306}}.
\newline\urlprefix\url{https://link.aps.org/doi/10.1103/PhysRevB.81.214306}

\bibitem{PhysRevB.84.144303}
M.~Haas, V.~Hizhnyakov, A.~Shelkan, M.~Klopov, A.~J. Sievers,
  \href{https://link.aps.org/doi/10.1103/PhysRevB.84.144303}{Prediction of
  high-frequency intrinsic localized modes in ni and nb}, Phys. Rev. B 84
  (2011) 144303.
\newblock \href {https://doi.org/10.1103/PhysRevB.84.144303}
  {\path{doi:10.1103/PhysRevB.84.144303}}.
\newline\urlprefix\url{https://link.aps.org/doi/10.1103/PhysRevB.84.144303}

\bibitem{khadeeva2011discrete}
L.~Z. Khadeeva, S.~V. Dmitriev, Y.~S. Kivshar,
  \href{https://doi.org/10.1134/S0021364011190106}{Discrete breathers in
  deformed graphene}, JETP letters 94~(7) (2011) 539--543.
\newblock \href {https://doi.org/10.1134/S0021364011190106}
  {\path{doi:10.1134/S0021364011190106}}.
\newline\urlprefix\url{https://doi.org/10.1134/S0021364011190106}

\bibitem{murzaev2017localized}
R.~T. Murzaev, D.~V. Bachurin, E.~A. Korznikova, S.~V. Dmitriev,
  \href{https://doi.org/10.1016/j.physleta.2017.01.014}{Localized vibrational
  modes in diamond}, Physics Letters A 381~(11) (2017) 1003--1008.
\newblock \href {https://doi.org/10.1016/j.physleta.2017.01.014}
  {\path{doi:10.1016/j.physleta.2017.01.014}}.
\newline\urlprefix\url{https://doi.org/10.1016/j.physleta.2017.01.014}

\bibitem{sato2007management}
M.~Sato, S.~Yasui, M.~Kimura, T.~Hikihara, A.~J. Sievers,
  \href{https://doi.org/10.1209/0295-5075/80/30002}{Management of localized
  energy in discrete nonlinear transmission lines}, EPL (Europhysics Letters)
  80~(3) (2007) 30002.
\newblock \href {https://doi.org/10.1209/0295-5075/80/30002}
  {\path{doi:10.1209/0295-5075/80/30002}}.
\newline\urlprefix\url{https://doi.org/10.1209/0295-5075/80/30002}

\bibitem{jin2014analysis}
L.~Jin, J.~Mei, L.~Li, \href{https://doi.org/10.1051/epjap/2014130565}{Analysis
  of intrinsic localised mode for a new energy harvesting cantilever array},
  The European Physical Journal Applied Physics 66~(1) (2014) 10902.
\newblock \href {https://doi.org/10.1051/epjap/2014130565}
  {\path{doi:10.1051/epjap/2014130565}}.
\newline\urlprefix\url{https://doi.org/10.1051/epjap/2014130565}

\bibitem{kimura2008stability}
M.~Kimura, T.~Hikihara,
  \href{https://doi.org/10.1016/j.physleta.2008.04.054}{Stability change of
  intrinsic localized mode in finite nonlinear coupled oscillators}, Physics
  Letters A 372~(25) (2008) 4592--4595.
\newblock \href {https://doi.org/10.1016/j.physleta.2008.04.054}
  {\path{doi:10.1016/j.physleta.2008.04.054}}.
\newline\urlprefix\url{https://doi.org/10.1016/j.physleta.2008.04.054}

\bibitem{kimura2009coupled}
M.~Kimura, T.~Hikihara,
  \href{https://doi.org/10.1016/j.physleta.2009.02.005}{Coupled cantilever
  array with tunable on-site nonlinearity and observation of localized
  oscillations}, Physics Letters A 373~(14) (2009) 1257--1260.
\newblock \href {https://doi.org/10.1016/j.physleta.2009.02.005}
  {\path{doi:10.1016/j.physleta.2009.02.005}}.
\newline\urlprefix\url{https://doi.org/10.1016/j.physleta.2009.02.005}

\bibitem{kimura2009capture}
M.~Kimura, T.~Hikihara, \href{https://doi.org/10.1063/1.3097068}{Capture and
  release of traveling intrinsic localized mode in coupled cantilever array},
  Chaos: An Interdisciplinary Journal of Nonlinear Science 19~(1) (2009)
  013138.
\newblock \href {https://doi.org/10.1063/1.3097068}
  {\path{doi:10.1063/1.3097068}}.
\newline\urlprefix\url{https://doi.org/10.1063/1.3097068}

\bibitem{cretegny19981d}
T.~Cretegny, S.~Aubry, S.~Flach, 1d phonon scattering by discrete breathers,
  Physica D: Nonlinear Phenomena 119~(1-2) (1998) 73--87.
\newblock \href {https://doi.org/10.1016/S0167-2789(98)00066-9,}
  {\path{doi:10.1016/S0167-2789(98)00066-9,}}.

\bibitem{hadipour2020interaction}
F.~Hadipour, D.~Saadatmand, M.~Ashhadi, A.~M. Marjaneh, I.~Evazzade, A.~Askari,
  S.~V. Dmitriev,
  \href{https://doi.org/10.1016/j.physleta.2019.126100}{Interaction of phonons
  with discrete breathers in one-dimensional chain with tunable type of
  anharmonicity}, Physics Letters A 384~(4) (2020) 126100.
\newblock \href {https://doi.org/10.1016/j.physleta.2019.126100}
  {\path{doi:10.1016/j.physleta.2019.126100}}.
\newline\urlprefix\url{https://doi.org/10.1016/j.physleta.2019.126100}

\bibitem{tsironis1999dependence}
G.~P. Tsironis, A.~R. Bishop, A.~V. Savin, A.~V. Zolotaryuk,
  \href{https://link.aps.org/doi/10.1103/PhysRevE.60.6610}{Dependence of
  thermal conductivity on discrete breathers in lattices}, Physical Review E
  60~(6) (1999) 6610--6613.
\newblock \href {https://doi.org/10.1103/PhysRevE.60.6610}
  {\path{doi:10.1103/PhysRevE.60.6610}}.
\newline\urlprefix\url{https://link.aps.org/doi/10.1103/PhysRevE.60.6610}

\bibitem{sato2006colloquium}
M.~Sato, B.~E. Hubbard, A.~J. Sievers,
  \href{https://link.aps.org/doi/10.1103/RevModPhys.78.137}{Colloquium:
  Nonlinear energy localization and its manipulation in micromechanical
  oscillator arrays}, Reviews of Modern Physics 78~(1) (2006) 137--157.
\newblock \href {https://doi.org/10.1103/RevModPhys.78.137}
  {\path{doi:10.1103/RevModPhys.78.137}}.
\newline\urlprefix\url{https://link.aps.org/doi/10.1103/RevModPhys.78.137}

\bibitem{sato2004optical}
M.~Sato, B.~E. Hubbard, A.~J. Sievers, B.~Ilic, H.~G. Craighead,
  \href{https://doi.org/10.1209/epl/i2003-10224-x}{Optical manipulation of
  intrinsic localized vibrational energy in cantilever arrays}, EPL
  (Europhysics Letters) 66~(3) (2004) 318--323.
\newblock \href {https://doi.org/10.1209/epl/i2003-10224-x}
  {\path{doi:10.1209/epl/i2003-10224-x}}.
\newline\urlprefix\url{https://doi.org/10.1209/epl/i2003-10224-x}

\bibitem{bang1996exploiting}
O.~Bang, P.~D. Miller, \href{https://doi.org/10.1364/OL.21.001105}{Exploiting
  discreteness for switching in waveguide arrays}, Optics letters 21~(15)
  (1996) 1105--1107.
\newblock \href {https://doi.org/10.1364/OL.21.001105}
  {\path{doi:10.1364/OL.21.001105}}.
\newline\urlprefix\url{https://doi.org/10.1364/OL.21.001105}

\bibitem{vicencio2003controlled}
R.~A. Vicencio, M.~I. Molina, Y.~S. Kivshar,
  \href{https://doi.org/10.1364/OL.28.001942}{Controlled switching of discrete
  solitons in waveguide arrays}, Optics letters 28~(20) (2003) 1942--1944.
\newblock \href {https://doi.org/10.1364/OL.28.001942}
  {\path{doi:10.1364/OL.28.001942}}.
\newline\urlprefix\url{https://doi.org/10.1364/OL.28.001942}

\bibitem{vicencio2004switching}
R.~A. Vicencio, M.~I. Molina, Y.~S. Kivshar,
  \href{https://doi.org/10.1103/PhysRevE.70.026602}{Switching of discrete
  optical solitons in engineered waveguide arrays}, Physical Review E 70~(2)
  (2004) 026602.
\newblock \href {https://doi.org/10.1103/PhysRevE.70.026602}
  {\path{doi:10.1103/PhysRevE.70.026602}}.
\newline\urlprefix\url{https://doi.org/10.1103/PhysRevE.70.026602}

\bibitem{kivshar1993peierls}
Y.~S. Kivshar, D.~K. Campbell,
  \href{https://doi.org/10.1103/PhysRevE.48.3077}{Peierls-nabarro potential
  barrier for highly localized nonlinear modes}, Physical Review E 48~(4)
  (1993) 3077.
\newblock \href {https://doi.org/10.1103/PhysRevE.48.3077}
  {\path{doi:10.1103/PhysRevE.48.3077}}.
\newline\urlprefix\url{https://doi.org/10.1103/PhysRevE.48.3077}

\bibitem{oster2003enhanced}
M.~J. M.~{\"O}ster, A.~Eriksson,
  \href{https://doi.org/10.1103/PhysRevE.67.056606}{Enhanced mobility of
  strongly localized modes in waveguide arrays by inversion of stability},
  Physical Review E 67~(5) (2003) 056606.
\newblock \href {https://doi.org/10.1103/PhysRevE.67.056606}
  {\path{doi:10.1103/PhysRevE.67.056606}}.
\newline\urlprefix\url{https://doi.org/10.1103/PhysRevE.67.056606}

\bibitem{hadvzievski2004power}
L.~Had{\v{z}}ievski, A.~Maluckov, M.~Stepi{\'c}, D.~Kip,
  \href{https://doi.org/10.1103/PhysRevLett.93.033901}{Power controlled soliton
  stability and steering in lattices with saturable nonlinearity}, Physical
  review letters 93~(3) (2004) 033901.
\newblock \href {https://doi.org/10.1103/PhysRevLett.93.033901}
  {\path{doi:10.1103/PhysRevLett.93.033901}}.
\newline\urlprefix\url{https://doi.org/10.1103/PhysRevLett.93.033901}

\bibitem{vicencio2006discrete}
R.~A. Vicencio, M.~Johansson,
  \href{https://doi.org/10.1103/PhysRevE.73.046602}{Discrete soliton mobility
  in two-dimensional waveguide arrays with saturable nonlinearity}, Physical
  Review E 73~(4) (2006) 046602.
\newblock \href {https://doi.org/10.1103/PhysRevE.73.046602}
  {\path{doi:10.1103/PhysRevE.73.046602}}.
\newline\urlprefix\url{https://doi.org/10.1103/PhysRevE.73.046602}

\bibitem{bang1995high}
O.~Bang, M.~Peyrard, High order breather solutions to a discrete nonlinear
  klein-gordon model, Physica D: Nonlinear Phenomena 81~(1-2) (1995) 9--22.

\bibitem{flach1998discrete}
S.~Flach, C.~R. Willis, Discrete breathers, Physics reports 295~(5) (1998)
  181--264.

\bibitem{maniadis2006mechanism}
P.~Maniadis, S.~Flach, Mechanism of discrete breather excitation in driven
  micro-mechanical cantilever arrays, Europhysics Letters 74~(3) (2006) 452.

\bibitem{page1990asymptotic}
J.~B. Page, \href{https://link.aps.org/doi/10.1103/PhysRevB.41.7835}{Asymptotic
  solutions for localized vibrational modes in strongly anharmonic periodic
  systems}, Phys. Rev. B 41~(11) (1990) 7835.
\newblock \href {https://doi.org/10.1103/PhysRevB.41.7835}
  {\path{doi:10.1103/PhysRevB.41.7835}}.
\newline\urlprefix\url{https://link.aps.org/doi/10.1103/PhysRevB.41.7835}

\bibitem{rosenberg1960nonlinear}
R.~M. Rosenberg, \href{https://doi.org/10.1115/1.3643948}{Normal modes of
  nonlinear dual-mode systems}, J. Appl. Mech. 27 (1960) 263--268.
\newblock \href {https://doi.org/10.1115/1.3643948}
  {\path{doi:10.1115/1.3643948}}.
\newline\urlprefix\url{https://doi.org/10.1115/1.3643948}

\bibitem{kerschen2009nonlinear}
G.~Kerschen, M.~Peeters, J.~C. Golinval, A.~F. Vakakis,
  \href{https://doi.org/10.1016/j.ymssp.2008.04.002}{Nonlinear normal modes,
  part {I}: {A} useful framework for the structural dynamicist}, Mechanical
  Systems and Signal Processing 23~(1) (2009) 170--194.
\newblock \href {https://doi.org/10.1016/j.ymssp.2008.04.002}
  {\path{doi:10.1016/j.ymssp.2008.04.002}}.
\newline\urlprefix\url{https://doi.org/10.1016/j.ymssp.2008.04.002}

\bibitem{peeters2009nonlinear}
M.~Peeters, R.~Vigui{\'e}, G.~S{\'e}randour, G.~Kerschen, J.~C. Golinval,
  \href{https://doi.org/10.1016/j.ymssp.2008.04.003}{Nonlinear normal modes,
  part {II}: toward a practical computation using numerical continuation
  techniques}, Mechanical systems and signal processing 23~(1) (2009) 195--216.
\newblock \href {https://doi.org/10.1016/j.ymssp.2008.04.003}
  {\path{doi:10.1016/j.ymssp.2008.04.003}}.
\newline\urlprefix\url{https://doi.org/10.1016/j.ymssp.2008.04.003}

\bibitem{marin1996breathers}
J.~L. Marin, S.~Aubry,
  \href{https://doi.org/10.1088/0951-7715/9/6/007}{Breathers in nonlinear
  lattices: Numerical calculation from the anticontinuous limit}, Nonlinearity
  9~(6) (1996) 1501--1528.
\newblock \href {https://doi.org/10.1088/0951-7715/9/6/007}
  {\path{doi:10.1088/0951-7715/9/6/007}}.
\newline\urlprefix\url{https://doi.org/10.1088/0951-7715/9/6/007}

\bibitem{saadatmand2018discrete}
D.~Saadatmand, D.~Xiong, V.~A. Kuzkin, A.~M. Krivtsov, A.~V. Savin, S.~V.
  Dmitriev, Discrete breathers assist energy transfer to ac-driven nonlinear
  chains, Physical Review E 97~(2) (2018) 022217.

\end{thebibliography}
\bibliographystyle{elsarticle-num}

\section*{Appendix A. Shift manipulation of ILM in Klein Gordon lattice with soft-type potential}
Shift manipulation of ILM in Klein Gordon lattice with soft-type potential is implemented with the same procedure as the manipulation in the lattice with hard-type anharmonicity. 
We take the potential expressed by the following equation
\begin{equation}
V(z)=\frac{z^2}{2}-\frac{z^4}{24}+\frac{z^6}{720}, 
\end{equation}
for the on-site potential with soft-type anharmonicity\ \cite{saadatmand2018discrete,hadipour2020interaction}. 
Thus, the equation of motion of the Klein Gordon lattice with dissipation and the periodic external driver is written as 
\begin{align}
m\ddot{u}_n+\gamma\dot{u}_n+u_n-\frac{u_n^3}{6}+\frac{u_n^5}{120}-C(u_{n-1}-2u_n&+u_{n+1}) \nonumber \\ 
&=F\cos\omega{t}, 
\end{align}
where the values of each parameter except $\omega$ are the same as in the lattice with hard-type anharmonicity. 
Here, we use the in-phase driver to sustain ILMs. 
The driving frequency $\omega$ is set at $0.7$, which is significantly smaller than the lowest band frequency ($\omega_{\mathrm{min}}=1$). 

\begin{figure}[htb]
     \begin{center}
    \includegraphics[width=85mm]{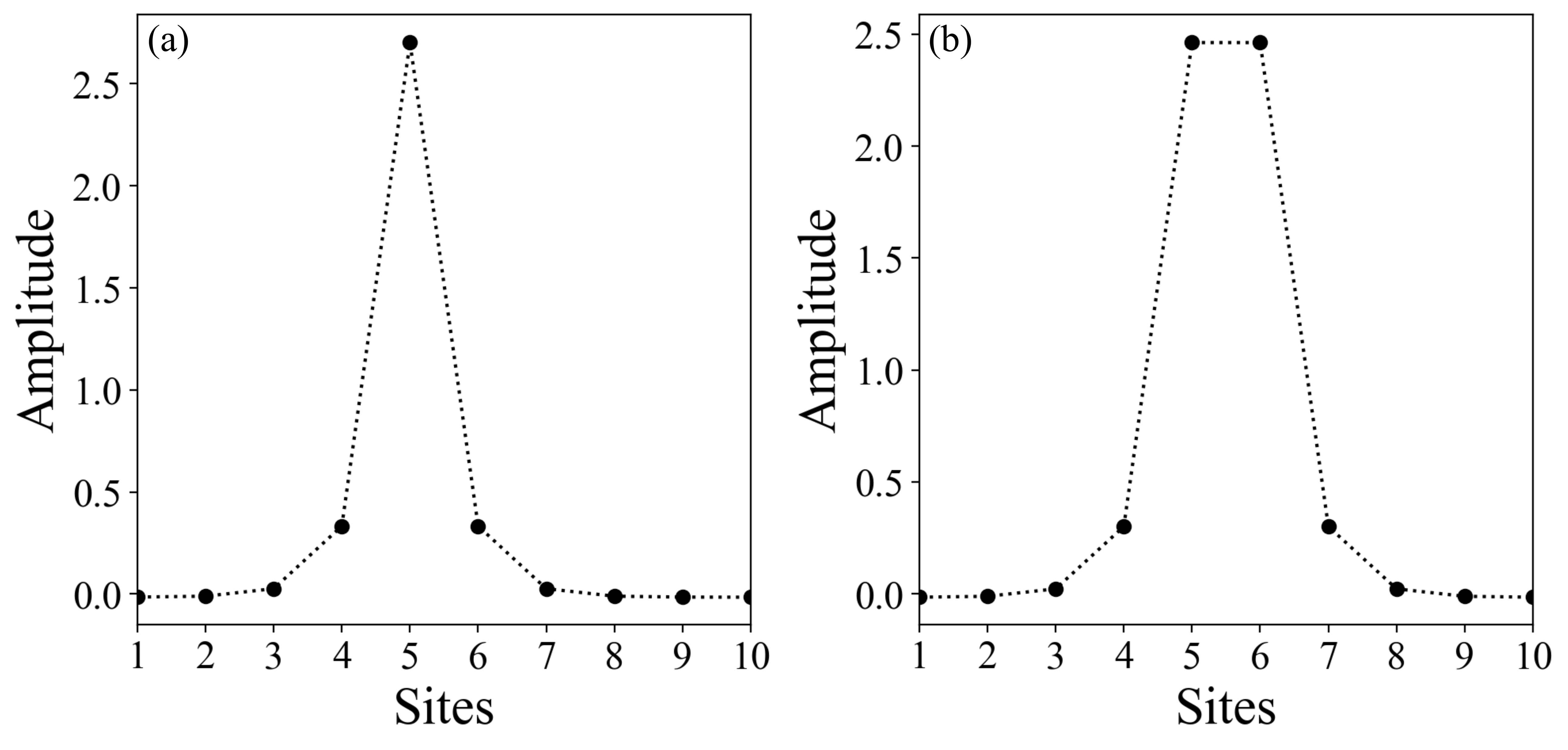}
    \end{center}
       \caption{Amplitude distribution of site centered mode (a) and bond centered mode (b). Each points indicate the position of particles when they have largest displacements. Each particles oscillate with the same period.}
     \label{fig:nine}
\end{figure}

There are two kinds of symmetrical ILM which are reverse-phase with the driver. 
One is a stable site centered mode, the other is an unstable bond centered mode. 
Fig. \ref{fig:nine}(a) shows the amplitude distribution of the site centered mode. 
Fig. \ref{fig:nine}(b) shows the amplitude distribution of the bond centered mode. 
Both of them has almost the same frequency as the driver and is reverse-phase to it. 
Thus, these ILMs are the strongly localized vibration modes with frequency far away from the phonon band. 

By reducing the ac driven Klein Gordon lattice to a two-degrees of freedom system, we obtain the equation of motion of the two-degree of freedom nonlinear system described by
\begin{align}
&\ddot{u}_{1}+\gamma\dot{u}_{1}+u_1-\frac{u_1^3}{6}+\frac{u_1^5}{120}-C(u_{2}-u_{1})=F\cos\omega{t}, \\
&\ddot{u}_{2}+\gamma\dot{u}_{2}+u_2-\frac{u_2^3}{6}+\frac{u_2^5}{120}-C(u_{1}-u_{2})=F\cos\omega{t},
\end{align}
where subscripts are replaced by 1 and 2 for simplicity. 

\begin{figure}[htb]
     \begin{center}
    \includegraphics[width=80mm]{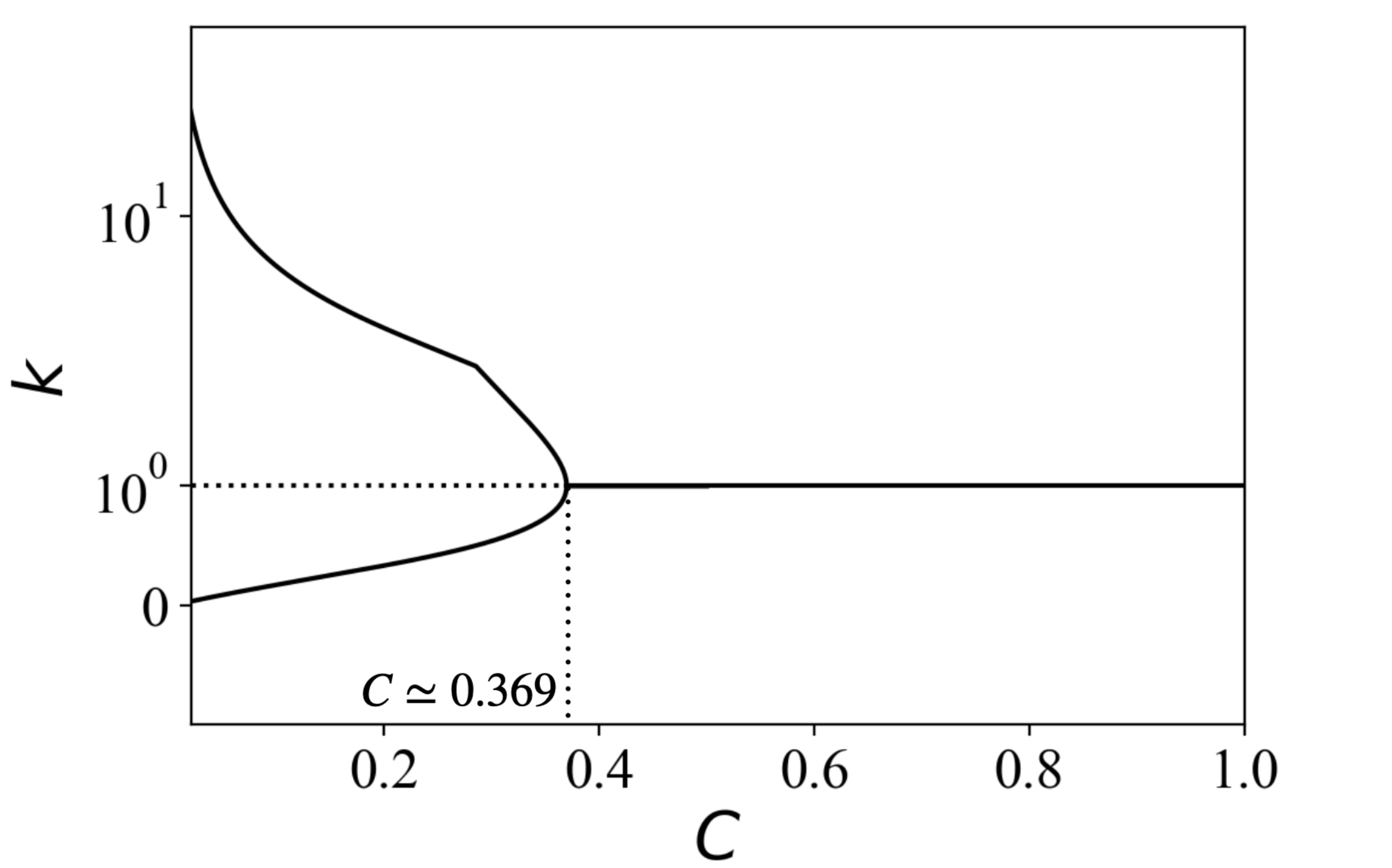}
    \end{center}
       \caption{Relationship between coupling constant and amplitude ratio $k$ in each of the coexisting NNMs. The solid line corresponds to stable NNMs. The dashed line corresponds to unstable NNMs.  } 
\label{fig:eleven}
\end{figure}

We find three coexisting NNMs in this system. 
These NNMs have reverse phase with the driver 
One is the in-phase mode with equal amplitude, which is unstable. 
This mode corresponds to unstable bond centered mode in the original system. 
The other two modes are localized modes which are the NNMs localized on each oscillator. 
These modes correspond to stable site centered mode. 
Fig. \ref{fig:eleven} shows amplitude ratio $k$ in each of the coexisting NNMs with respect to the coupling constant $C$. 
The solid line indicates stable NNMs and the dashed line unstable NNMs. 
The straight line at k=1 corresponds to the in-phase mode. 
The curve above the bifurcation corresponds to NNMs localized at oscillator 1. 
The curve below the bifurcation point to NNMs localized at oscillator 2.
When the coupling constant is large, only stable anti-phase mode exists. 
As the value of $C$ is decreased, the in-phase mode loses its stability at $C\simeq0.369$ and two stable coexisting localized modes appear. 
Thus, the pitchfork bifurcation of NNMs also appears in the two-degree of freedom system with soft-type anharmonicity. 

Based on this bifurcation diagram, we implement the switching of two coexisting localized mode by appropriately tuning the coupling constant. 
We change the coupling constant $C(t)$ adiabatically according to Eq. (\ref{eq:couplingconstant}) with $C_{\mathrm{b}}=0.369$. 
Now, the driver period $T=2\pi/\omega$ is more than twice larger than in the case of the system with hard-type anharmonicity. 
Accordingly, we need to set the value of $T_{\mathrm{I}}$ and $T_{\mathrm{R}}$ appropriately. 

\begin{figure*}[htb]
     \begin{center}
    \includegraphics[width=182mm]{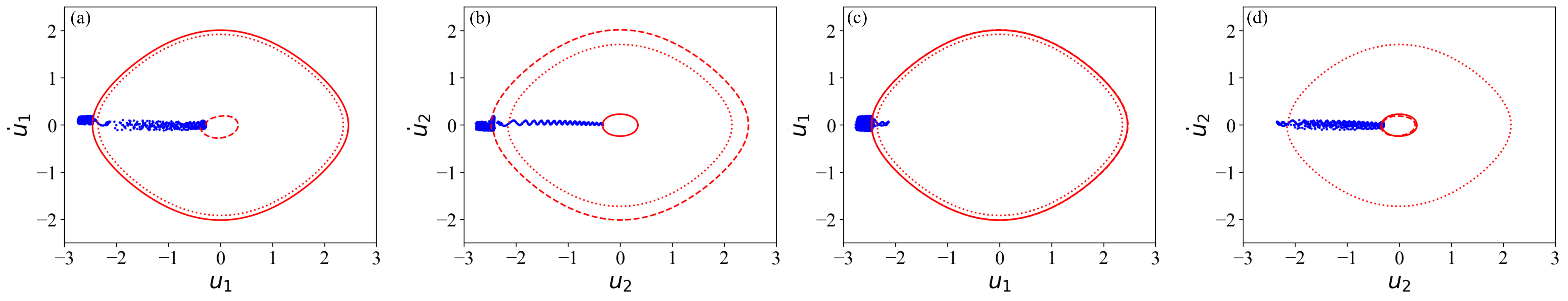}
    \end{center}
       \caption{(a) and (b) Simulation results of the switching when $T_{\mathrm{I}},\ T_{\mathrm{C}}$, and $T_{\mathrm{R}}$ are set at $250T,\ 20T$, and $250T$ respectively. (c) and (d) Simulation results of the switching when $T_{\mathrm{I}},\ T_{\mathrm{C}}$, and $T_{\mathrm{R}}$ are set at $250T,\ 30T$, and $250T$ respectively. Points (Blue) indicates Poincare map of orbits of each oscillators during the manipulation. Solid line is trajectory for a period of each oscillators before the switching and dotted line is during period $T_\mathrm{C}$, and dashed line is after the switching.}
     \label{fig:twelve}
\end{figure*}

Figs. \ref{fig:twelve}(a) and \ref{fig:twelve}(b) show the simulation results when $T_{\mathrm{I}},\ T_{\mathrm{C}}$, and $T_{\mathrm{R}}$ are set at $250T,\ 20T$, and $250T$, respectively. 
Points indicate Poincare maps and lines each oscillator trajectory for one period before, during, and after the switching. 
Before the switching, the oscillator 1 has a large amplitude while the oscillator 2 has a small amplitude. 
During period $T_\mathrm{C}$, the oscillators have almost the same amplitude. 
After the switching, the oscillator 1 has a small amplitude while the oscillator 2 has a large amplitude. 
We can summarize the result as follows. 
First, during the period $T_\mathrm{I}$, the NNM localized at oscillator 1 (the solid line) gradually approaches the in-phase mode (the dotted line). 
The in-phase mode appears during the period $T_{\mathrm{C}}$. 
Finally, during the period $T_\mathrm{R}$, the in-phase mode approaches NNM localized at oscillator 2 (the dashed line). 
Thus, under this condition, the switching is achieved between the coexisting localized modes. 

Figs. \ref{fig:twelve}(c) and \ref{fig:twelve}(d) show the simulation results when $T_{\mathrm{I}},\ T_{\mathrm{C}}$, and $T_{\mathrm{R}}$ are set at $250T,\ 30T$, and $250T$, respectively. 
Before the switching, the oscillator 1 has a large amplitude while the oscillator 2 has a small amplitude. 
During period $T_\mathrm{C}$, the oscillators have almost the same amplitude. 
After the switching, the oscillator 1 has a large amplitude while the oscillator 2 has a small amplitude. 
We can summarize the result as follows. 
First, during the period $T_\mathrm{I}$, the NNM localized at oscillator 1 (the solid line) gradually approaches the in-phase mode (the dotted line). 
The in-phase mode appears during the period $T_{\mathrm{C}}$. 
Finally, during the period $T_\mathrm{R}$, the in-phase mode approaches the original state (the dashed line). 
Therefore, under this condition, we will not achieve the switching of coexisting localized modes. 
These showed that the switching depends on the way of tuning the coupling constant as in the case of hard-type anharmonicity. 

\begin{figure}[htb]
     \begin{center}
    \includegraphics[width=88mm]{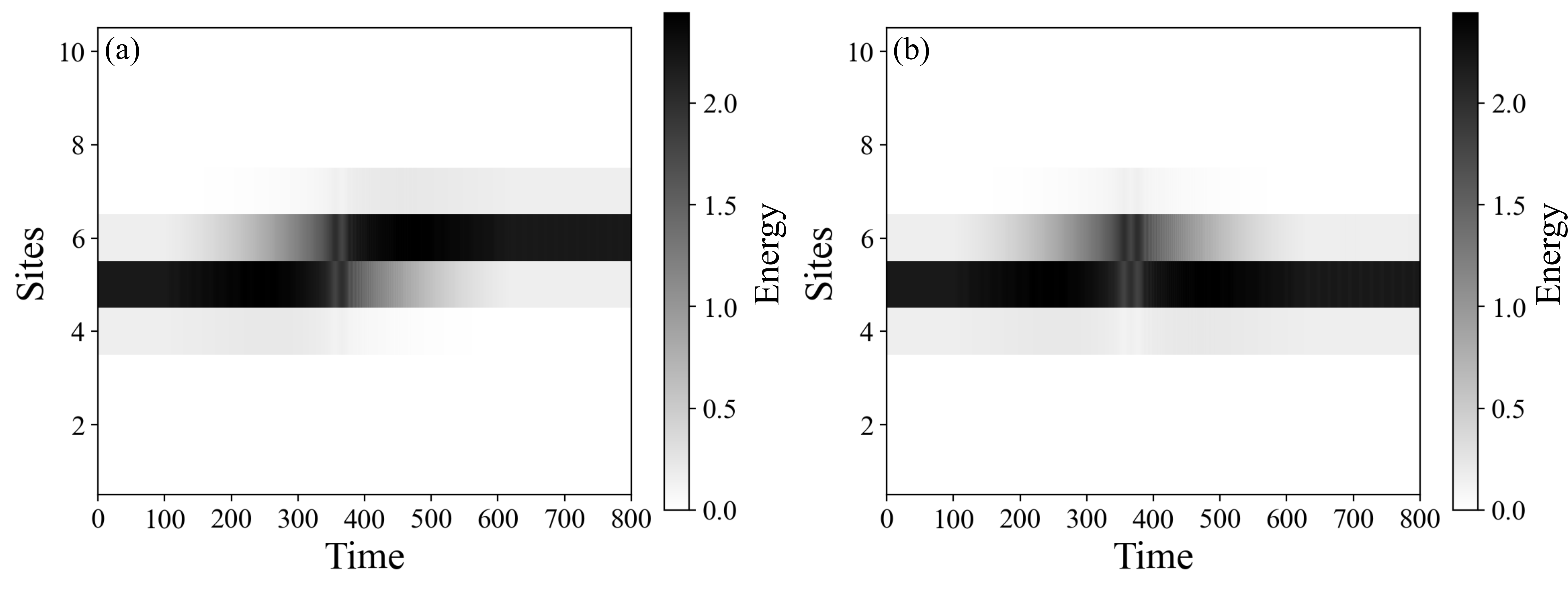}
    \end{center}
       \caption{Time development of energy of each sites during the shift manipulation of ILM in the lattice with soft-type anharmonicity. The unit time on the horizontal axis is $T$ which is the period of driver. (a) Simulation results when $T_{\mathrm{I}},\ T_{\mathrm{C}}$, and $T_{\mathrm{R}}$ are set at $250T,\ 20T$, and $250T$ respectively. (b) Simulation results when $T_{\mathrm{I}},\ T_{\mathrm{C}}$, and $T_{\mathrm{R}}$ are set at $250T,\ 30T$, and $250T$ respectively. }
     \label{fig:fourteen}
\end{figure}

Based on the results of the reduced system, we move an ILM from the 5th site to the 6th site. 
Thus, we change a coupling constant between the 5th and the 6th site adiabatically as in the case of switching between coexisting localized modes. 
Fig. \ref{fig:fourteen}(a) shows the behavior of ILM during the manipulation when $T_{\mathrm{I}},\ T_{\mathrm{C}}$, and $T_{\mathrm{D}}$ are set at $250T,\ 20T$, and $250T$, respectively. 
In this case, the manipulation succeeds. 
Fig. \ref{fig:fourteen}(b) shows the behavior of ILM during the manipulation when $T_{\mathrm{I}},\ T_{\mathrm{C}}$, and $T_{\mathrm{D}}$ are set at $250T,\ 30T$, and $250T$, respectively. 
In this case, the manipulation fails. 
These results indicate that the manipulation depends on the way of tuning the coupling constant.
Further investigation of the dependency on the process of the manipulation reveals that two different intervals for the value of $T_{\mathrm{C}}$ appear alternately: the manipulation succeeds in one interval and fails in the other.   
Near the boundaries between these intervals, fine recursive structures appear.
These results are similar to those in the lattice with hard-type anharmonicity. 
Consequently, for the accurate manipulation of ILM, we need to set the value of $T_{\mathrm{C}}$ appropriately. 

\section*{Appendix B. The numerical method for obtaining NNMs and ILMs}
We describe the numerical method for finding particular oscillation solutions such as NNMs and ILMs with the period $T$ in each system. 
We rewrite the equation of motion of the systems into the following state space form
\begin{equation}
\dot{\bm{x}}=\bm{f}(\bm{x},\ t)
\end{equation}
where $\bm{x}=(\bm{u,\ \dot{\bm{u}}})$ is the $2n$-dimensional state space and $\bm{f}(\bm{x},\ t)$ is a function of period $T$ with respect to $t$. 
The numerical solution of this dynamical system for initial conditions $\bm{x}(0)=\bm{x}_0$ is written as $\bm{x}(t)=\phi(t,\ \bm{x}_0)$. 

We then define the map that numerically integrates the dynamical system from a given initial state $\bm{x}_0$ over the period $T$ by 
\begin{equation}
F_T(\bm{x}_0)=\phi(T,\ \bm{x}_0)
\end{equation}
where $F_T$ depends parametrically on $T$ and on the initial condition $\bm{x}_0$. 
If $\bm{x}_{p}$ is the solution such as NNMs and ILMs with the period $T$, $F_T(\bm{x}_p)=\bm{x}_p$ holds. 
Therefore, we can obtain these solutions by numerically solving the two-point boundary-value problem defined by the following periodicity condition
\begin{equation}
G_T({\bm{x}_{p}})=F_T(\bm{x}_p)-\bm{x}_p=0
\end{equation}
where $G_T(\bm{x}_0)=F_T(\bm{x}_0)-\bm{x}_0$ represents the difference between the initial condition and the state of the system integrated for $T$ from the initial conditions. 

The shooting method is the most popular numerical technique to solve this problem. 
In this case, since $T$ is fixed, the method consists in finding initial conditions iteratively by the Newton-Raphson algorithm.
The Newton–Raphson algorithm is used to correct an initial guess and to converge to the actual solution.
The Newton–Raphson iteration scheme generates the following map
\begin{align}
\bm{x}_{n+1}&=\bm{x}_n-[\partial _{\bm{x}_0}G_T(\bm{x}_n)]^{-1}G_T(\bm{x}_n) \\ \nonumber
&=\bm{x}_n-[\partial_{\bm{x}_0}F_T-I]^{-1}G_T(\bm{x}_n)
\end{align}
where $\partial_{\bm{x}_0}$ represents the derivative with respect to initial conditions $\bm{x}_0$ and $\partial_{\bm{x}_0}F$ is the monodromy matrix. 
This map converges if an initial guess is sufficiently close to the solution. 
To obtain a good initial guess, we use the continuation method or consider a limiting case of the system. 
\end{document}